\documentclass[twocolumn,showpacs,preprintnumbers,amsmath,amssymb]{revtex4}

\usepackage[unicode]{hyperref}
\usepackage[caption=false]{subfig}
\usepackage{graphics}
\usepackage{epstopdf}
\usepackage{epsfig}
\usepackage{graphicx}
\graphicspath{ {./images/} }
\usepackage{dcolumn}
\usepackage{bm}
\usepackage[caption=false]{subfig}

\begin{document}

\title{ Explosive percolation on scale-free multifractal weighted planar stochastic lattice
}%

\author{M. K Hassan and M. M. Rahman 
}%
\date{\today}%

\affiliation{
University of Dhaka, Department of Physics, Theoretical Physics Group, Dhaka 1000, Bangladesh \\
}

\begin{abstract}

In this article, we investigate explosive bond percolation (EBP) with product rule, formally known as Achlioptas process, 
on a scale-free multifractal weighted planar stochastic lattice (WPSL). One of the key features of the EBP transition
is the delay, compared to corresponding random bond percolation (RBP), in the onset of spanning cluster. However, when it happens, it happens so
dramatically that initially it was believed, albeit ultimately proved wrong, 
that explosive percolation (EP) exhibits first order transition. In the case of EP, much efforts were devoted to resolving the issue of its order
of transition and almost no effort being devoted to find critical point, critical exponents etc., to classify 
it into universality classes. This is in sharp
contrast to the classical random percolation. We do not even know all the exponents of EP
for regular planar lattice or for Erd\"{o}s-Renyi network. We first find numerically the critical point $p_c$ 
and then obtain all the critical exponents $\beta, \gamma, \nu$ as well as the Fisher exponent 
$\tau$ and the fractal dimension $d_f$ of the spanning cluster. We also compare our results for EBP with those of the RBP and find that
all the exponents of EBP obeys the same scaling relations as do the RBP. Our findings
suggests that EBP is no special except the fact that the exponent $\beta$ is unusually
small compared to that of RBP.
\end{abstract}


\maketitle

\section{Introduction}

The idea of percolation was first conceived by Flory in 1941 in the context of gelation transition \cite{ref.Flory}. However, percolation as a mathematical model
was first formulated by Broadbent and Hammersley in 1957 to understand the motion of gas molecules through the maze of pores in carbon granules 
filling a gas mask  \cite{ref.Broadbent}. Since then, it remained one of the most studied 
theories in statistical physics. To study percolation one has to first choose a skeleton, an empty lattice 
or a graph, which has two entities namely sites or nodes and bonds or links. 
One of its entity, depending on whether it is bond or site type percolation, is occupied with probability $p$ 
independent of the state of its neighbors \citep{ref.Stauffer}. 
As the occupation probability $p$ is tuned starting from $p=0$ clusters, 
i.e. contiguous occupied sites, are gradually formed, merged and grown. Remarkably, in the process there 
appears a cluster that spans across 
the entire linear size of the lattice at a certain non-trivial threshold value $p_c$. When it happens, it happens so abruptly 
that many observable quantities 
diverge at $p_c$. This is reminiscent of continuous thermal phase transition where physical 
properties like susceptibility, specific heat etc. diverge in a similar fashion \cite{ref.Stanley}. 
Phase transitions are classified according to how an order parameter (OP), 
a quantity which is zero in one phase and non-zero in the other, varies in the 
immediate vicinity of the critical point. For instance, phase transitions are called 
discontinuous (or first order) if the 
OP itself is discontinuous at $p_c$ and they are called continuous (or second order) if OP is continuous across 
the whole range of $p$. 
Percolation transition is well-known as a paradigmatic model of second order phase transition since the OP, 
the relative size of the spanning
cluster $P$, grows from zero at $p_c$ following a power-law $P\sim (p-p_c)^\beta$ which is exactly how
magnetization behaves in ferromagnetic transition.
The insights into the percolation theory therefore facilitates 
the understanding of phase transition and critical phenomena which
is one of the most elegant field of research in statistical and condensed matter physics
\cite{ref.Schwabl}.

In 2009 Achlioptas {\it et al.}  proposed 
a biased occupation rule, known as the Achlioptas process (AP), 
that encourages slower growth of the larger clusters and faster growth of the smaller clusters instead of random 
occupation in classical percolation \cite{ref.Achlioptas}. According to this rule
a pair of bonds are first picked uniformly at random from all possible distinct 
links. However, of the two,
only the one that satisfies the pre-selected rule is finally chosen to occupy and the other one is discarded.
The preset rule is usually chosen so that it discourages the growth of the larger clusters
and encourages the growth of the smaller clusters. As a result, the percolation
threshold is delayed and hence the corresponding $p_c$ is always higher than the 
case where only one bond is always selected. 
Furthermore, it is natural to expect that close to $p_c$ nearly equal 
sized clusters, waiting to merge, are so great in number that occupation of a 
few bonds results in an abrupt global connection and thus the name ``Explosive Percolation'' (EP).
Through their seminal paper Achlioptas {\it et al.} claimed for the first time that EP 
can describe the first order phase transition (see for recent reviews in \cite{ref.saberi, ref.review}). Their results jolted the scientific community through 
a series of claims, unclaims and counter-claims \cite{ref.Ziff_1, ref.Ziff_2, ref.Radicchi, ref.Souza,
 ref.Cho_1, ref.Cho_2, ref.Raissa, ref.Hayasaka, ref.Costa_1, ref.Costa_2} .
However, recent studies on the percolation transition under original AP rule and its various variants suggest 
that the transition is actually continuous in character 
\cite{ref.Riordan, ref.Grassberger}. Moreover, there are also claims that albeit it is
a continuous transition it also exhibits some unusual behaviors 
\cite{ref.Ziff_2, ref.Cho_2, ref.Grassberger, ref.Tian, ref.Bastas}. For instance, the critical
exponent $\beta$ of the order parameter is so small in comparison to that of its random percolation that it can be easily mistaken for zero
which can lead to conclude OP suffering a jump \cite{ref.da_Costa}. The EP model was first
implemented on the ER network. The idea was then extended to other planar
lattices and to scale-free networks \cite{ref.Ziff_1, ref.Choi}. As many variants
of the EP model were introduced, it became more apparent that EP actually describes continuous phase transition. 
Recently, it has been further generalized by picking
a fixed $m\geq 2$ number of candidate bonds at each step instead of a pair bonds only. It has been
claimed that AP in the limit $m\rightarrow \infty$ on a lattice can still yield a discontinuous percolation
transition at $p_c$ \cite{ref.Qian}.

It is noteworthy to mention that the extent of connectivity, which depends on the distance of its state from the $p_c$, is highly important 
in many systems . 
There are systems where large-scale connectivity is desired and there are systems where it can be a liability too. 
For instance, in the case of virus spreading on social or computer network, 
a higher $p_c$ is desired so that even if $p$ is high the spread of viruses can still be contained in small, 
isolated clusters. 
However, in the case of communication network, a smaller $p_c$ is desired so that the system can have large 
scale connectivity even at small $p$. 
Note that the smaller the $p_c$, the better the connectivity even at small $p$. The flexibility in controlling 
the location of the percolation 
threshold $p_c$ therefore can be of great interest.
One of the advantages of the EP model is that we can either enhance or lessen the $p_c$ value
simply by inverting the condition of the AP rule. We can also tune the $p_c$ by using various variants 
of the AP rule. Besides, finding the critical exponents of
the EP model can also be of significant interest since most of the studies on EP have primarily been  focused 
on resolving
the debate whether it describes continuous or discontinuous transition. This is
in sharp contrast to random percolation (RP) for which we know critical exponents for a wide range of regular and random lattices
( see Refs. \cite{ref.Wiki, ref.Hsu} and references there in). 
One of the extraordinary findings of RP is that the critical
exponents are found to be universal in the sense that they depend only on the dimension of the
lattice. That is, regardless of whether the skeleton is a square, triangular or
honeycomb lattice as long as they are planar in the sense that their dimension coincides with the dimension 
of the space where they are embedded, they will share the same critical exponents regardless of whether 
the percolation is of bond or site type. However, recently we have performed site and bond RP on a 
multifractal scale-free weighted planar stochastic lattice (WPSL) and found an exception for the 
first time \cite{ref.Hassan_Rahman_1,ref.Hassan_Rahman_2}. To be precise, we found that both site and bond 
percolation on WPSL belong to the same universality class which is different from the universality 
class where all the known planar latices belong.
It is note worthy to mention that there have not been enough efforts to classify the EP into universality classes since most studies on 
EP focused on resolving the issue whether it describes continuous or discontinuous phase transition.
Having just overcome that transient phase, it is  
now time to focus on finding critical exponents for various lattices or graphs to classify them 
into universality classes.

The focus of this article is on finding the critical exponents of explosive bond percolation (EBP) 
on WPSL and compare its results
with those of the random bond percolation (RBP) on the same lattice.
It is a special lattice with some unique features that no other known lattice
has. For instance on one hand, unlike network or graph, it has
property of lattice as its sites are spatially embedded. On the other, unlike lattices, its dual display the 
property of networks as its coordination number distribution follows a power-law. Besides, unlike regular lattice, the
size of its cells are not equal rather the distribution of the area size of its blocks obeys dynamic
scaling \cite{ref.Hassan_Dayeen}. Moreover, the dynamics of the growth of this lattice is governed by 
infinitely many conservation
laws, one of which being the trivial conservation of total area. One more interesting property of the WPSL is
that each of the non-trivial conservation law can be used as a multifractal measure and hence it is also 
a multi-multifractal \cite{ref.Hassan_Dayeen}. Krapivsky and Ben-Naim also showed that it exhibits multiscaling \cite{ref.Krapivsky_Naim}.
Yet another property of the WPSL is that it can be mapped as a network if we consider the center of each block
as a node and the common border between block as the link between the center of the corresponding nodes. Interestingly,
the degree distribution of the corresponding network exhibits power-law \cite{ref.Hassan_Pavel}. 
Considering these links as bonds we perform percolation on the WPSL and find numerically the values of the critical exponents $\beta, \gamma, \nu$ as well as
the exponent $\tau$ that characterizes the cluster size distribution function $n_s(p_c)$ and the fractal dimension $d_f$
that characterizes the spanning cluster. One of the advantage that WPSL has over network or graph
is that we can identify the spanning cluster. Note that networks or graphs do not have
edges, sides or boundaries and hence the relative size of the largest cluster is defined as the order
parameter instead that of the spanning cluster. We compare the results of the EP of bond with those of the RP and 
found a distinct set of exponents.
In particular, we find that the exponent $\beta$ of EP is remarkably smaller than that of the RP on the WPSL which justifies
the name explosive. We also show that the scaling functions of the EP are different from those of the RP on the WPSL.
To the best of our knowledge, this is the first comprehensive study of EP where all the usual
critical exponents are obtained. We show that these values satisfy all the scaling and hyperscaling relations
among themselves like they do in the case of RP. To test our values, we further use the idea of data-collapse which
stands as an ultimate test of their accuracy. These result reveals that EP model is just another variants of percolation
theory.

The rest of this article is organized as follows. In section II we briefly discuss the construction and 
the properties of the WPSL.
In section III, we first find the percolation threshold $p_c$ for EP using the idea of spanning probability $W(p)$
that there is a cluster that spans across the entire lattice at $p$.  
Second, using the same $W(p)$ we also find an estimate for the critical exponent $\nu$.
Third, we use the idea of percolation probability (order parameter), ratio of the size of the spanning cluster to the
size of the lattice, and the idea of mean cluster size to find the numerical estimates
for the critical exponent $\beta$ and $\gamma$ respectively. Besides,
we find the exponents $\tau$ and $d_f$ of the cluster size distribution function $n_s(p)$ and spanning
cluster at $p_c$. Finally in section IV we summarize our findings.


\section{WPSL and its properties}

We start by giving a brief description of how we construct the WPSL \cite{ref.Hassan_Pavel}.
 It starts with a square of unit
area which we regard as an initiator. The generator then divides the initiator, in the first step, randomly with
uniform probability into four smaller blocks. In the second step and thereafter, the generator
 is applied to only one of the blocks. The question is: How do we pick
that block when there are more than one blocks? The most generic choice would be to pick preferentially 
according to their areas so that the higher the area the higher the probability to be picked. For instance,
in step one, the generator divides the initiator 
randomly into four smaller blocks. Let us label their areas  starting from the top
left corner and moving clockwise as $a_1,a_2,a_3$ and $a_4$. 
But of course the way we label is totally arbitrary and will bear no
consequence to the final results of any observable quantities. Note that $a_i$ is the area of the $i$th block
which can be well regarded as the probability of picking the $i$th block. Interestingly, 
these probabilities are naturally normalized $\sum_i a_i=1$
since we choose the area of the initiator equal to one. In step two, we pick one of the four blocks preferentially
with respect to their areas. Consider that we pick the block $3$ and apply the generator onto it 
to divide it randomly into four smaller blocks. Thus the label $3$ is now redundant and hence we recycle
it to label the top left corner while the rest of three new blocks are labelled  $a_5, a_6$ and $a_7$ in a clockwise
fashion. In general, in the $j$th step, we pick one out of $3j-2$ blocks 
 preferentially with respect to area and divide randomly into four blocks. The detailed algorithm
can be found in Ref. \cite{ref.Hassan_Dayeen, ref.Hassan_Pavel}.

\begin{figure}
\includegraphics[width=6.5cm,height=6.0cm,clip=true]{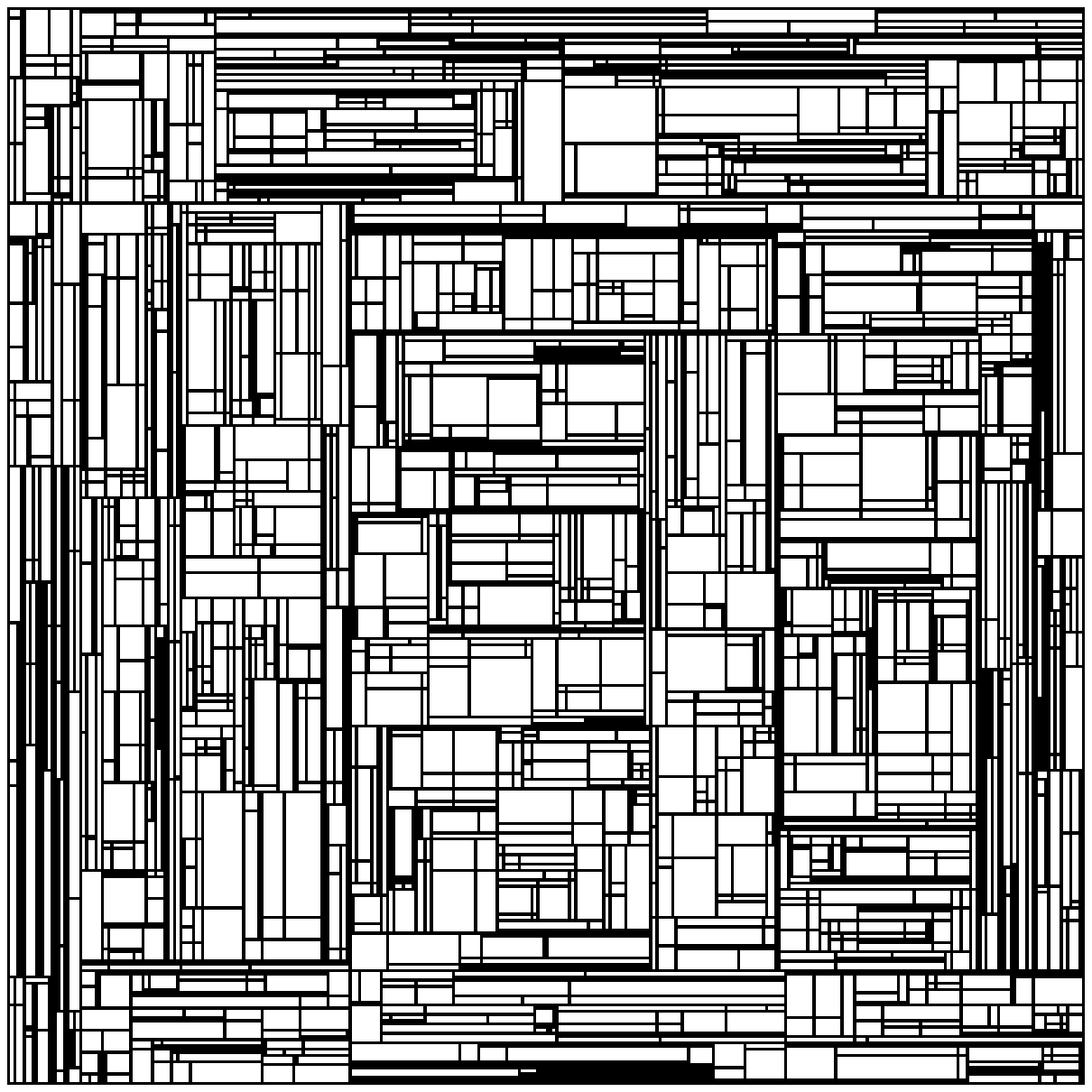}
\caption{A snapshot of the weighted planar stochastic lattice.
}
\label{fig:1}
\end{figure}%

The creation of the WPSL can also describe the following processes. First, two mutually perpendicular
cuts grow upon random sequential nucleation  of a seed in the initiator. Second, the tips of the two cuts 
move  with  a  constant velocity until
they  are hit or intercepted either by another cut or by the  boundary. The algorithm can also describe kinetics of fragmentation of planar 
objects through the effect the effects of size and shape can be dealt in a minimalist way \cite{ref.Hassan_Rodgers, ref.redner_book}.
Despite the simplicity, the process 
yet yields a lattice that looks seemingly complex, manifestly intricate 
and inextricably intertwined, which makes it an interesting candidate to look if there are some scaling and 
order (geometrical or topological).
Perhaps a representative snapshot of the lattice, see Fig. \ref{fig:1}, can give a better impression about 
the lattice than a mere description. In this work we shall treat this as a random lattice which has the following non-trivial properties.

\begin{itemize}

\item 
One of the interesting observable physical quantities for the WPSL can well be the block size
 distribution function $C(a,t)$ where $a$ represents the area of the blocks. 
It describes the concentration of blocks of the area within the size range $a$ and $a+da$ at time $t$. 
We have recently shown that it exhibits dynamic scaling \cite{ref.Hassan_Dayeen}. Note that the WPSL is a disordered lattice that emerges through evolution and hence
it can only be useful if 
the snapshots taken at different late stages are similar. In physics, similarity and self-similarity have a 
specific meaning.  
Two snapshots of the WPSL taken at two very different times can be similar if they differ in the 
numerical values of the dimensional quantities while
the numerical values of the corresponding dimensionless quantities coincide.

\item The dynamics of the process is governed by infinitely many non-trivial conservation laws including the trivial 
conservation of the total area of all the blocks of the lattice. That is,
if the $i$th block is described by the size of its length $x_i$ and width $y_i$ then we find that the numerical 
value of the quantity $M_m=\sum_i^N x_i^{(4/m)-1}y_i^{m-1}$ remains 
the same regardless of the size of the lattice for any value of $m$ where $m=2$ corresponds to the total area (the trivial conservation laws).

\item Each of the nontrivial conservation law $M_m$ are distributed in the WPSL such that the fraction of this quantity that the $i$th block 
has is $p_i\sim x_i^{(4/m)-1}y_i^{m-1}$. After constructing the partition function, the $q$th moment of 
$p_i$, and measuring that with a square of side $\delta$ equal to the mean block area 
of the WPSL we find it follows a power law with exponent $\tau(q,m)=(1-q)D_q(m)$. The Legendre transform of $\tau(q,m)$ gives the multifractal $f(\alpha)$ spectrum revealing 
that each non-trivial conserved quantity is a multifractal measure and hence the WPSL is a multi-multifractal \cite{ ref.Hassan_Pavel}.

\item We can map the WPSL as a network if we regard each block of the WPSL as node and the common border between blocks as links. We find that
the fraction of the total nodes (blocks) which has degree $k$,
that describes the probability $P(k)$ that a node picked at random has a degree $k$, 
is known as the degree distribution $P(k)$. This is equivalent
to the coordination number distribution in the WPSL. We find that $P(k)$ decays obeying a power-law \cite{ref.Hassan_Dayeen, ref.Hassan_Pavel}.
Thus we see that the WPSL in one hand has the properties of networks since unlike a typical lattice
its coordination number distribution follow power-law. On the other hand, unlike networks, its nodes are
spatially embedded and have edges or boundaries.

\item  It has a mixture of properties of both lattice and graph.
In one hand, like lattice, its cells are embedded spatially in the space of dimension $D=2$ and on the other, 
like scale-free network, its coordination number distribution follows a power-law.

\item It also has interesting neighborhood statistics. For instance, the mean area $\langle A\rangle_k$ of only those blocks which share $k$ neighbours obeys Lewis law,
i.e., $\langle A\rangle_k \propto k$, for up to $k=8$ and beyond that it reaches to a constant exponentially.  
Besides, if we regard  $m_k$ as the mean or typical number of neighbors of only those blocks which has
exactly $k$ neighbours then we find that $km_k$ is a constant (statistical sense). It implies
that the Aboav-Weaire law, $km_k\propto k$, is violated in the WPSL for the entire range of $k$ \cite{ref.Hassan_Dayeen}.

\end{itemize}

\section{Explosive bond percolation on the WPSL}

In this article, we investigate explosive and random bond percolation on the WPSL. 
In either case we have to first understand
what is bond in the context of WPSL. Second, how many bonds are there in the WPSL of size 
$N$ blocks. To understand what is bond in the WPSL we first map the WPSL into a network which we
call dual of the WPSL. This is obtained by replacing each block by a node at their centers and the common
border between two blocks by a link connecting the corresponding nodes.
It is note worthy to mention that the network corresponding to the dual of the WPSL 
has surface sites while  networks
or graphs do not have such surface sites. It is this feature of the network corresponding to dual of the
WPSL which gives spanning probability a meaning in this case. 
As regard to the second question, the number of bonds in the WPSL of fixed size varies in each
independent realization. Interestingly, its average over many $\cal{N}$ independent realizations reaches a constant value 
as we let $\cal{N}\rightarrow \infty$. Initially, the dual of the WPSL consisting of $N$ nodes (blocks)
has exactly $N$ number of cluster of size one. We first label all the bonds as $1,2,...,m$ so that a bond
$e_{mn}$ connects two sites $m$ and $n$ which belong to clusters say of sizes $s_m$ and $s_n$ respectively.
Then, in the explosive bond percolation (EPB) according to the AP rule, we pick
a pair of bonds $e_{ij}$ and $e_{kl}$ at random from all possible distinct 
bonds. However, this is only a trial attempt from which one of the links that minimizes the product 
of the size of the two clusters to
which it attaches is finally occupied and the attempt to occupy the other is discarded. That is, if 
$s_is_j<s_ks_l$ then the bond $e_{ij}$ is occupied and if $s_ks_l<s_is_j$ then $e_{kl}$ is occupied
while attempt to occupy $e_{kl}$ is discarded in the former case and $e_{ij}$ in the latter.
On the other hand, in the random bond percolation (RBP) only one bond is 
picked at random and occupied regardless of the size of the component clusters it attaches.
In either case, each time we occupy a bond, a 
cluster at least of size two or more is formed. The size of the cluster in the case
bond percolation, be it EBP or RBP, is measured by the number
of sites connected by occupied bonds. Understanding the nature of percolation transition and accurately 
predicting the percolation threshold are of fundamental importance
and it is one of the central tasks in the study of percolation \citep{ref.ziff_pc_1, ref.ziff_pc_2}. 
On the other hand, it is thought that if finding the critical exponents in random percolation is hard then finding them in the explosive percolation is even harder
especially the $\beta$ value. In this article we will find all the critical exponents and verify them using the scaling and hyperscaling relations.

\subsection{Spanning probability $W(p)$}

\begin{figure}
\centering
\subfloat[]
{
\includegraphics[height=4.0 cm, width=2.4 cm, clip=true,angle=-90]
{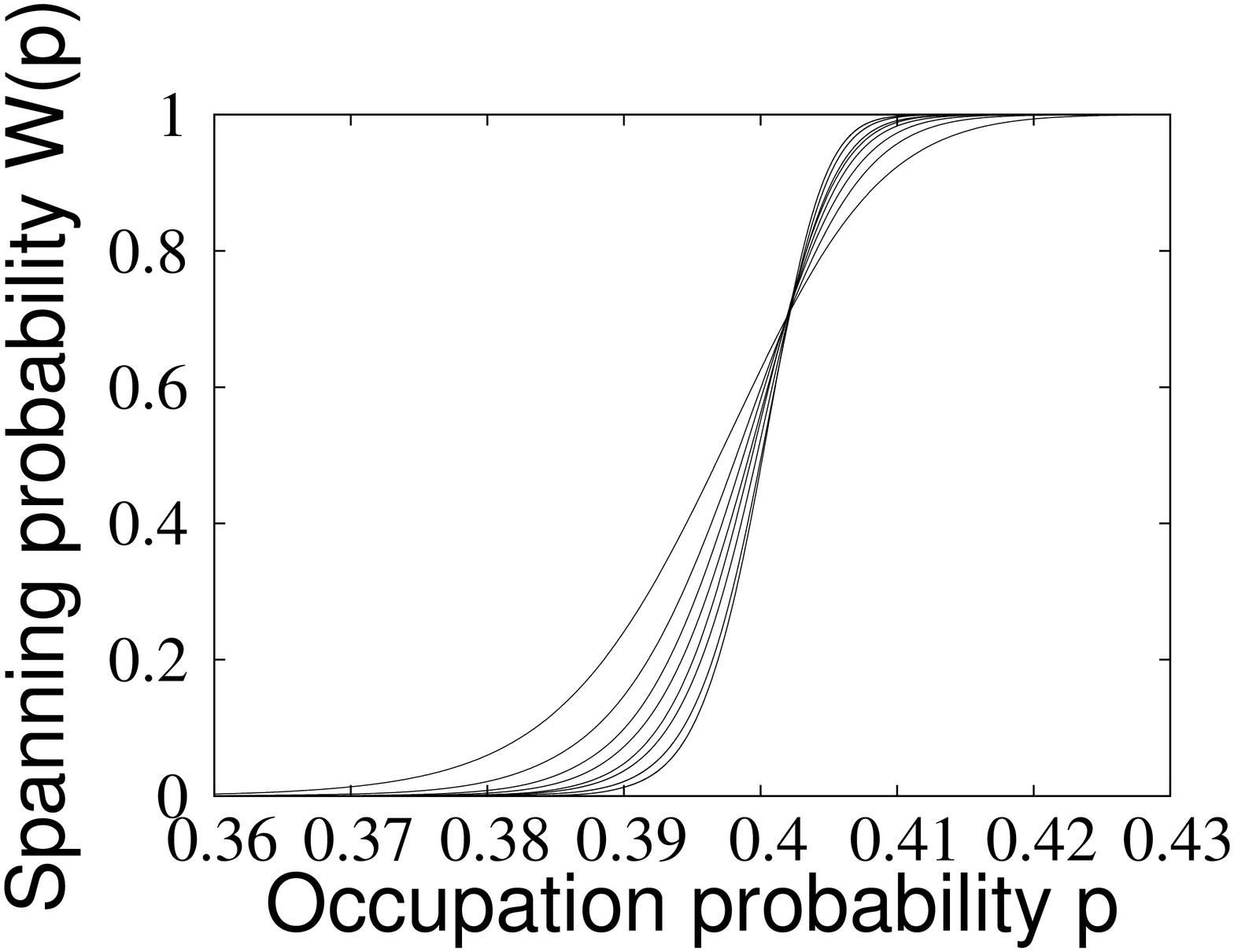}
\label{fig:2a}
}
\subfloat[]
{
\includegraphics[height=4.0 cm, width=2.4 cm, clip=true, angle=-90]
{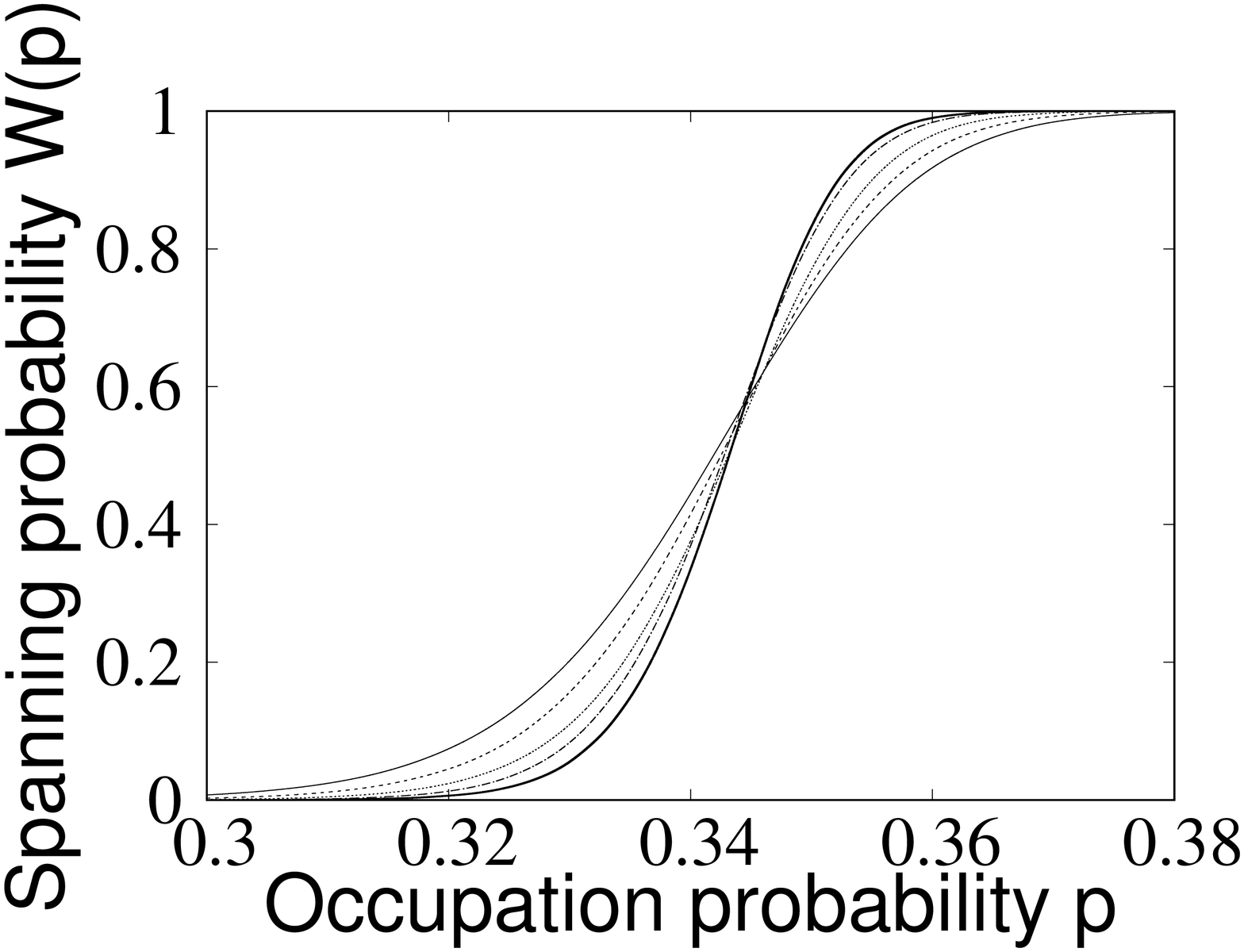}
\label{fig:2b}
}

\subfloat[]
{
\includegraphics[height=4.0 cm, width=2.4 cm, clip=true, angle=-90]
{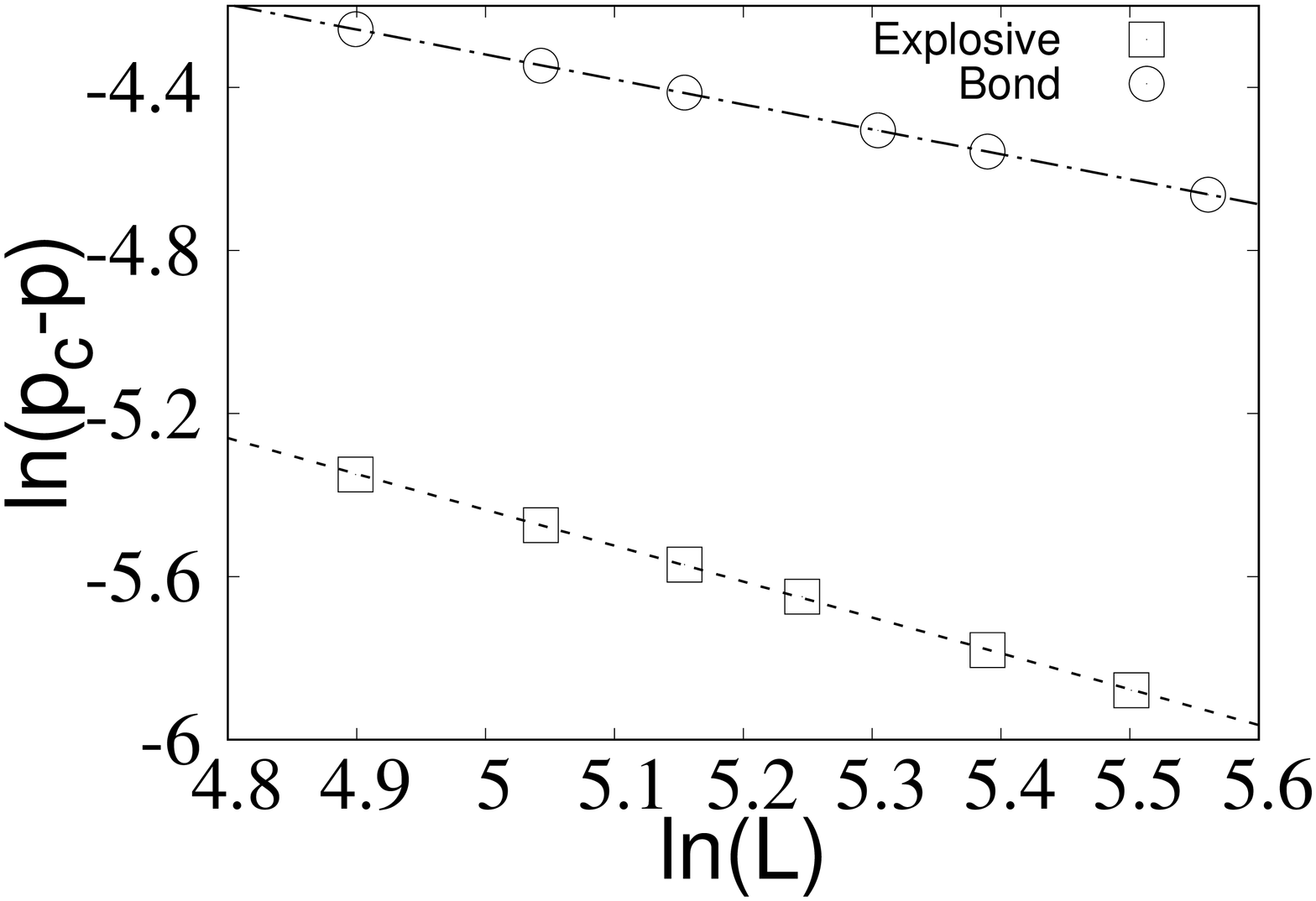}
\label{fig:2c}
}
\subfloat[]
{
\includegraphics[height=4.0 cm, width=2.4 cm, clip=true, angle=-90]
{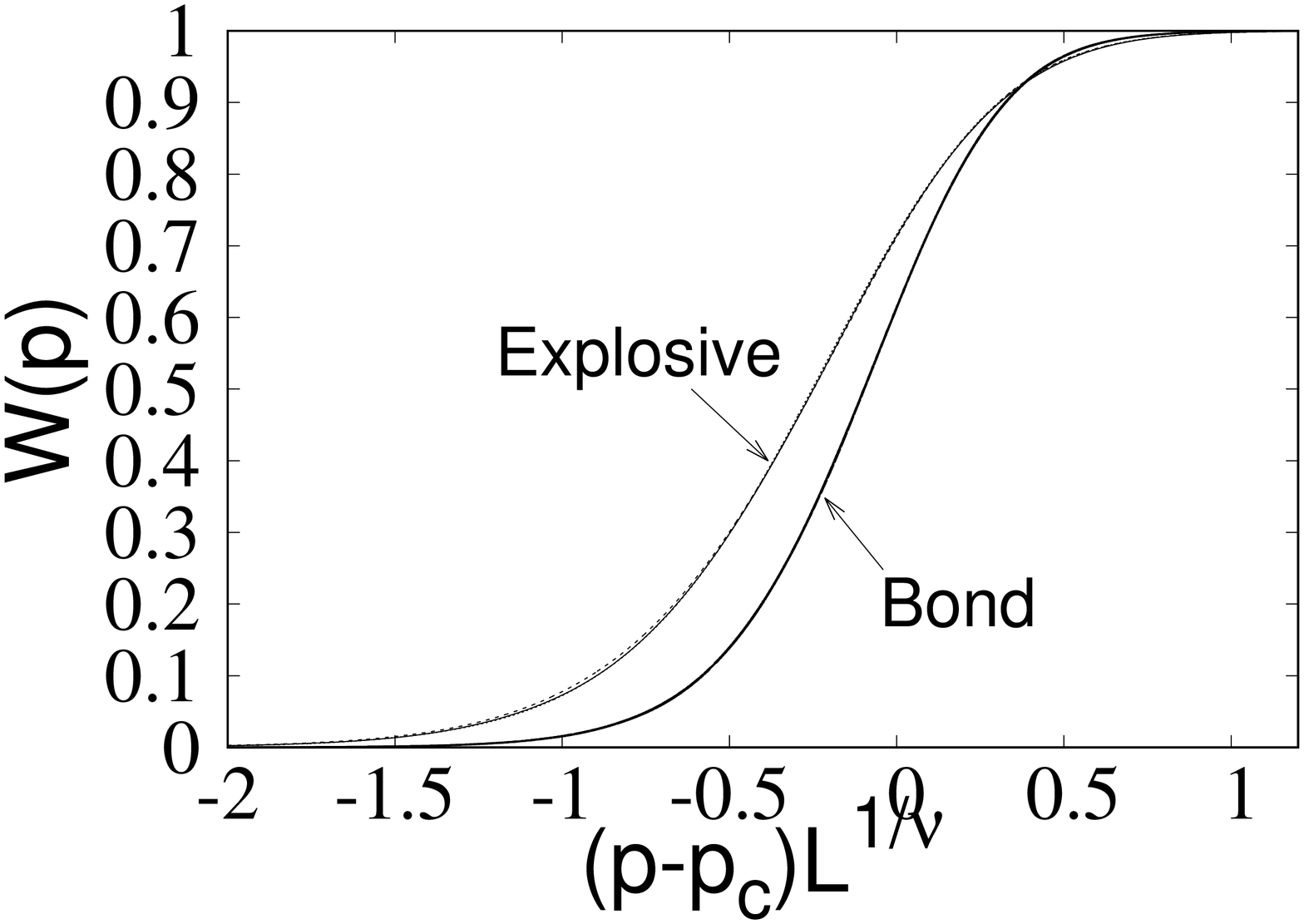}
\label{fig:2d}
}

\caption{Spanning probability $W(p,L)$ vs $p$ in WPSL for (a) explosive bond and 
(b) random bond percolation. The simulation result of the percolation threshold is
$p_{c}=0.4021$ for EBP and $0.3457$ for random bond percolation. In (c) we plot $\log(p-p_c)$ vs $\log L$ 
for both the cases. The two lines have slopes $1/\nu=0.8801 \pm 0.0049$ and $0.6117\pm 0.0074$ 
for explosive and random bond respectively. In (d) we plot dimensionless quantities $W$ vs $(p-p_c)L^{1/\nu}$ 
and we find distinct plots in (a) and (b) collapse superbly into their own scaling function.
} 
\label{fig:2abcd}
\end{figure}

We first attempt to find the percolation threshold $p_c$ and the critical exponent $\nu$ for explosive percolation. 
The best observable quantity to find both is the spanning probability $W(p)$.
The spanning probability $W(p)$ describes the likelihood of finding a 
cluster that spans across the entire system
either horizontally or vertically at the occupation probability $p$. To find how $W(p)$
behaves with the control parameter $p$ we perform many, say $M$, independent realizations 
under the same identical conditions. In each realization for a given finite system size we take record of the 
$p_c$ value at which the spanning cluster appears for the first time. Thus there is a spanning cluster for all $p>p_c$ and hence we set zero for all $p<p_c$
and one for each $p\geq p_c$ value. We then count all the $1$s whose sum can at best be equal to $M$ where $M$ is the number of independent realizations.
To find a regularity or a pattern
among all the $M$ numbers for a given $p$ value we count all the ones at which a spanning cluster exists. We use this data
to obtain the relative frequency of occurrence at a given $p$ that we regard
as the spanning probability $W(p)$. In Figs. \ref{fig:2a} and \ref{fig:2b}, we show a set of plots of $W(p)$ for explosive and random bond percolation 
respectively as a function of
$p$ where distinct curves represent different system size $L=\sqrt{N}$. One of the significant features of such plots is that all 
the distinct plots for different size $L$ meet at one particular $p$ value. Each curve represents a polynomial 
equation in $p$ for a given $L$. 
The significance of the meeting is that 
it is the root of all the polynomial equations and it is actually the critical point $p_c$.  In the case of 
explosive bond percolation, we find $p_c=0.4021$ which
is higher than  $p_c=0.3457$ for random bond percolation as expected since the AP rule systematically 
delays the emergence of spanning cluster. We can further tune the value of $p_c$ to get 
a better estimate using the finite-size scaling for $W(p)$.

It is interesting to note that the idea of spanning probability can also be used to find the critical exponent 
$\nu$. In pursuit of this we find it worthwhile to observe
the direction of shift of the $W(p)$ vs $p$ curves on either side of $p_c$ as the system size $L$ increases. 
This shift shows a clear sign of march of the curves towards $p_c$ from either side revealing that 
$W(p)$ will ultimately be like a step function in the limit $L\rightarrow \infty$. 
In other words it is expected that $W(p)=0$ for $p\leq p_c$ and $W(p)=1$ for $p>p_c$ which is the
 hallmark of percolation transition. We can quantify the extent at which they are marching by
measuring the magnitude of the difference $(p_c- p)$ for different $L$ for a fixed $W(p)$ value. We
do it by drawing a horizontal line at a given value of $W$, preferably at the position where this difference is
the most to minimize the error, and take records of the difference $p_c-p$ as a function of system size $L$.
Plotting the resulting data after taking log of both the variables we
find a straight line, see Fig. \ref{fig:2c},  with slope $0.8801\pm 0.0049$ for explosive and $0.6135 
\pm 0.0038$ random bond percolation. The slopes are actually equal to inverse of $\nu$ and hence we can write
\begin{equation}
\label{eq:10}
p_c- p\sim L^{-{{1}\over{\nu}}}.
\end{equation}
Indeed, it implies that in the limit $L\rightarrow \infty$ all the $p$ take the
value $p_c$ revealing that $W(p)$ will ultimately become a step function. To further test and to 
further verify the value of $\nu$ we use the finite-size scaling
hypothesis
\begin{equation}
\label{eq:fss1}
 W(p,L)=L^{-a/\nu}\phi_nu((p-p_c)L^{1/\nu}). 
\end{equation}
Now $W(p)$ is a step function means $a=0$ and hence if we plot $W(p)$ vs $(p_c- p)L^{{{1}\over{\nu}}}$ then
 all the distinct plots of $W(p)$ vs $p$ should
collapse in to a single universal curve. Indeed, we find an excellent collapse of all the distinct plots 
for different sizes which is shown in Fig. \ref{fig:2d}.
The quality of data-collapse also provides a test that the values of $p_c$ and $\nu$ obtained numerically are quite accurate up to an excellent extent.

\subsection{Order parameter: Percolation probability $P(p)$}

\begin{figure}
\centering
\subfloat[]
{
\includegraphics[height=4.0 cm, width=2.4 cm, clip=true,angle=-90]
{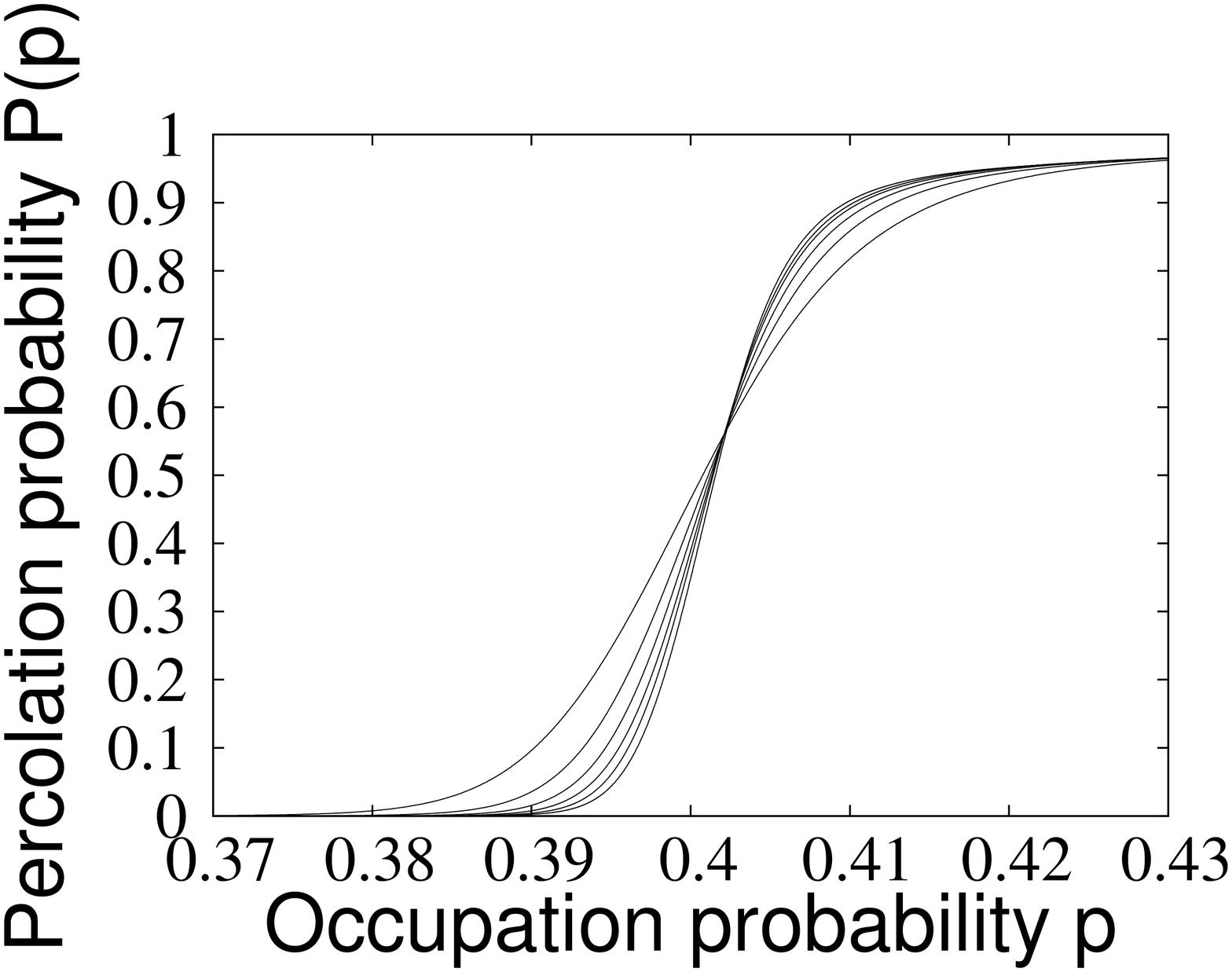}
\label{fig:3a}
}
\subfloat[]
{
\includegraphics[height=4.0 cm, width=2.4 cm, clip=true, angle=-90]
{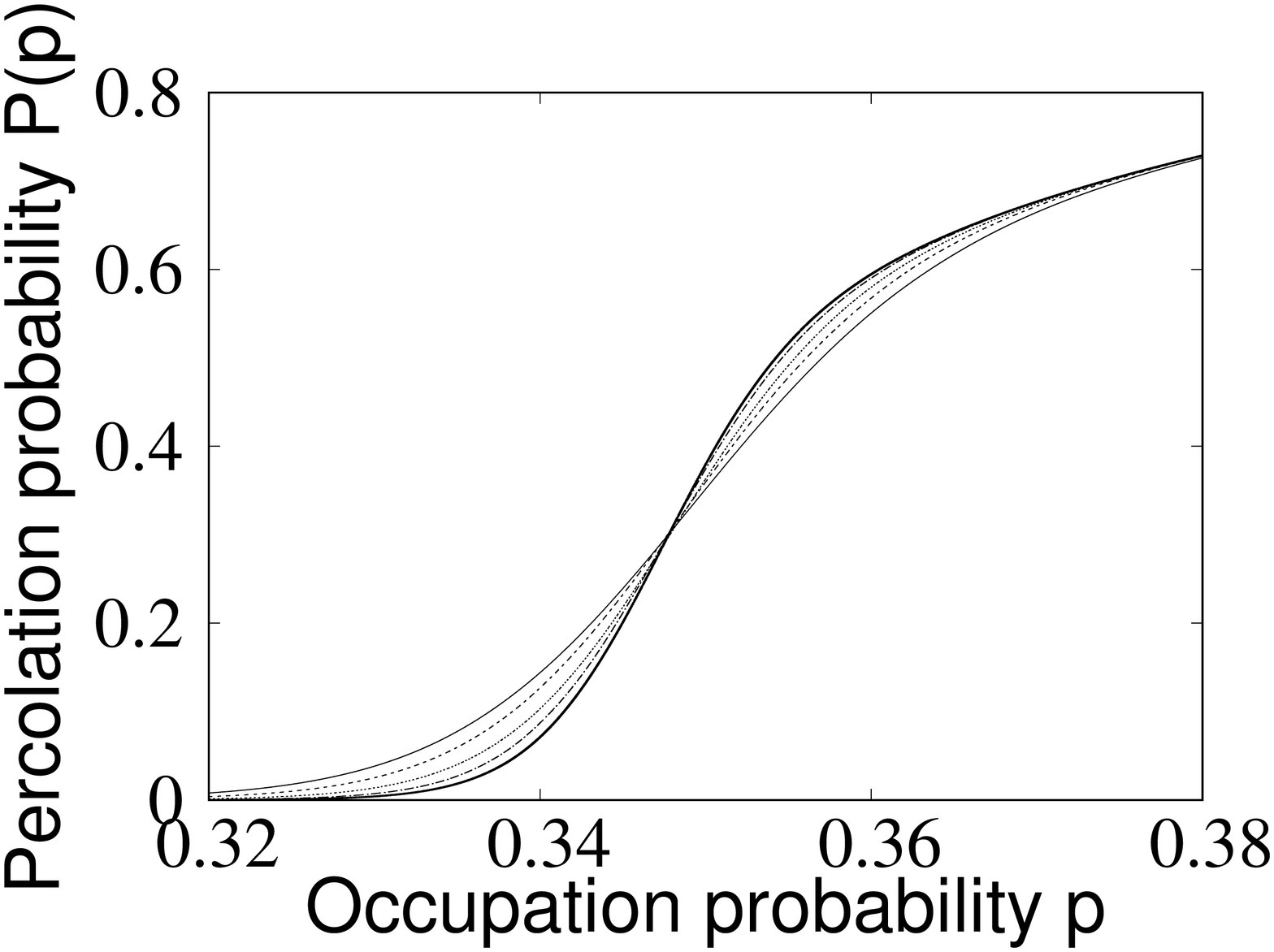}
\label{fig:3b}
}

\subfloat[]
{
\includegraphics[height=4.0 cm, width=2.4 cm, clip=true, angle=-90]
{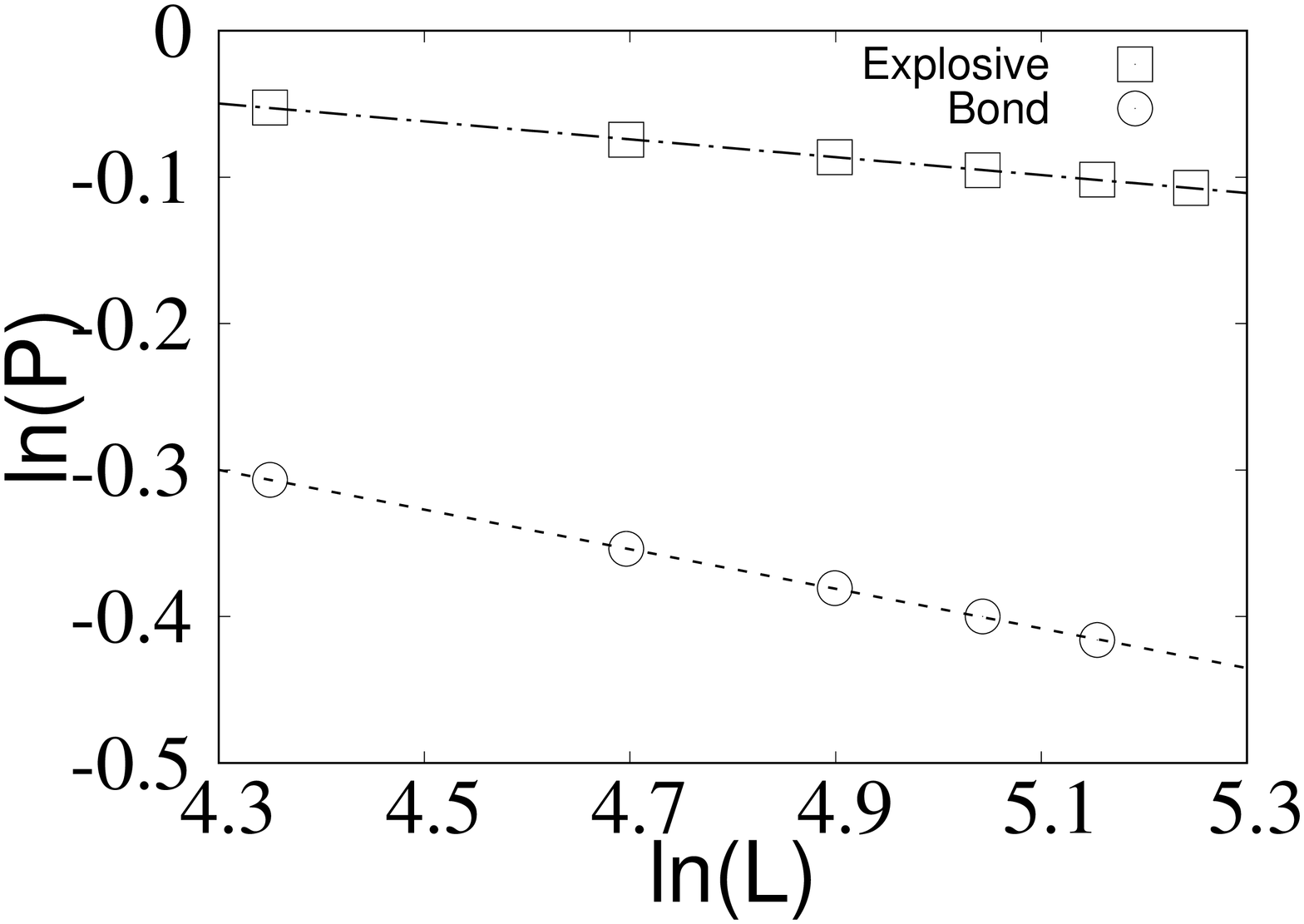}
\label{fig:3c}
}
\subfloat[]
{
\includegraphics[height=4.0 cm, width=2.4 cm, clip=true, angle=-90]
{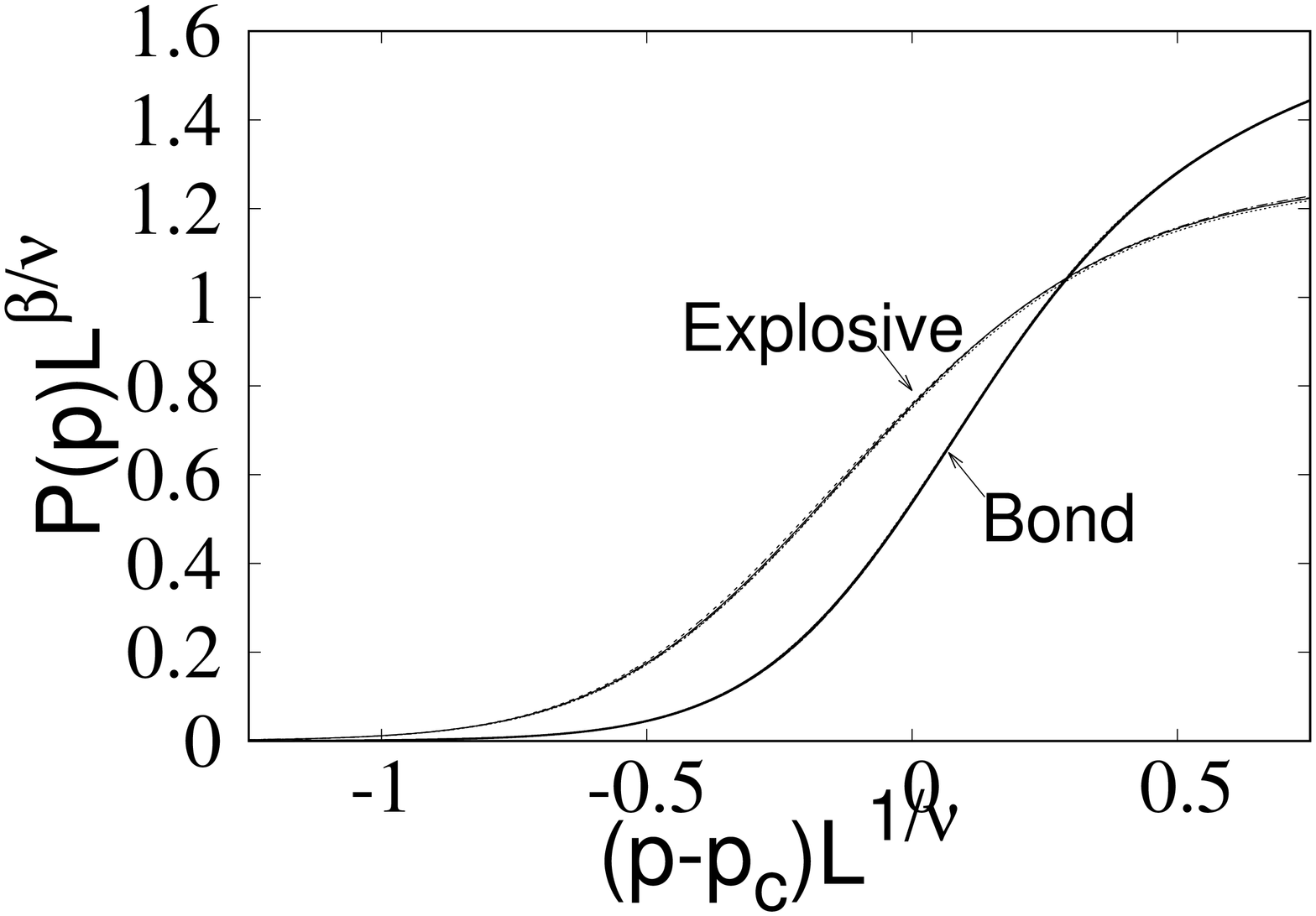}
\label{fig:3d}
}

\caption{Percolation strength or percolation probability $P(p,L)$ in WPSL for (a) explosive bond and (b) random 
bond percolation.
In (c) we plot $\log P$ vs $\log L$ using data for fixed value of $(p-p_c)L^{1/\nu}$ and find almost
parallel lines with slopes $\beta/\nu=0.059835 \pm 0.00038$ and $ 0.1357 \pm 0.0002$ for random bond
percolation respectively which clearly implies that the critical exponent $\beta=0.0679$ is negligibly 
small for explosive compare to $\beta=0.222$ random bond on the same lattice. In (d)  
we plot $PL^{\beta/\nu}$ vs $(p-p_c)L^{1/\nu}$ and find distinct plots of (a) and (b) 
collapses into their own scaling functions.
} 
\label{fig:3abcd}
\end{figure}

To find the critical exponent $\beta$ we have to consider the equivalent counterpart of the order parameter in percolation. 
In percolation, the percolation probability $P(p)$ (also
sometimes called percolation strength) is defined as ratio of the size spanning cluster $A_{{\rm span}}$ to the size of the largest possible cluster $N$
(which is actually the size of the lattice). In the case of percolation on graph or network we, however, use the largest cluster $A_{{\rm largest}}$ 
in place of the spanning cluster since in
network the term spanning does not exist. In the present case, we can use the former since we can recognize the spanning cluster in the WPSL. 
We plot percolation probability $P$ in Figs. \ref{fig:3a} and \ref{fig:3b} as a function of $p$ for both explosive 
and random bond percolation respectively. Looking at the plots, one
may think that all the plots for different $L$ meet at a single unique point like it does for $W(p)$ vs $p$ plot. 
However, if one zooms in, then it becomes 
apparent it is not so and hence the $p_c$ value from this plot will not be as accurate as it is from  the 
$W(p)$ vs $p$ plot. 
We also find that $P(p)$ is not 
strictly equal to zero at $p<p_c$, rather there is always a non-zero chance of finding a spanning cluster
even at $p<p_c$ as long as the system size $L$ is finite. However, the plots of $P$ vs $p$ also give a clear 
indication that the chances of getting 
spanning cluster at $p<p_c$ diminishes with increasing $L$. There is also a lateral shift of the $P$ vs $p$ plot 
to the left for $p>p_c$. The extent of this shift, however, decreases but never becomes a step function 
like in the case of $W(p)$ vs $p$ plot.
In contrast to RBP, the rise of $P$ in EBP is much sharper. However, the growth of $P(p)$ for EBP in the Erd\"{o}s-Renyi network is so sharp that it can be mistaken
as a step function in which case the critical exponent would have been zero. It has been later found that $\beta$ value in that case is actually too low. 
One of the goals of this work is to find the $\beta$ value for explosive percolation on the WPSL and compare its value with that RBP.

To show that the percolation probability $P$ does not suffer a jump or discontinuity we need to show
it behaves like $P\sim (p-p_c)^\beta$ with $\beta>0$ since $\beta=0$ would mean first order transition. In order
 to check if the exponent $\beta=0$ or $\beta>0$ for infinite system size $L$, we again apply the idea of finite-size scaling
\begin{equation}
\label{eq:fss2}
P(p,L)\sim L^{-\beta/\nu}\phi_\beta((p-p_c)L^{1/\nu}).
\end{equation}
We already know the $\nu$ value from the $W(p)$ vs $p$ curves. 
To find $\beta/\nu$ we first plot $P(p)$ vs $(p_c- p_c(L))L^{{{1}\over{\nu}}}$
and find that unlike $W(p)$ vs $(p_c- p_c(L))L^{{{1}\over{\nu}}}$ it does not collapse.
It immediately implies that $\beta\neq 0$ and hence $P(p)$ does not suffer a jump revealing that EP is not first order. 
To find the value of $\beta/\nu$, we measure the heights $P_{{\rm height}}$  at a given value of $(p-p_c)L^{1/\nu}$
for different $L$. We then plot $\log (P_{{\rm height}})$ vs $\log (L)$ as shown in Fig. \ref{fig:3c}
and find a straight line with slopes $\beta/\nu=0.0598 \pm 0.0003$ for EBP
and $ 0.1357 \pm 0.0002$ for RBP revealing that
\begin{equation}
\label{eq:11}
P(p,L)\sim L^{-\beta/\nu}.
\end{equation}
Now according to Eq. \ref{eq:fss2} if we now plot $PL^{\beta/\nu}$ vs $(p-p_c)L^{1/\nu}$ all the distinct plots of 
$P$ vs $p$ should collapse into a single universal curve.
Indeed, we see  that all the distinct plots of Figs. \ref{fig:3a} and \ref{fig:3b} collapse superbly into 
their own universal scaling curves (see Fig. \ref{fig:3d}).
Now using Eq. (\ref{eq:10}) in Eq. (\ref{eq:11}) to eliminate $L$ in favor of $p-p_c$  we get
\begin{equation}
\label{eq:pp4}
P\sim (p-p_c)^\beta,
\end{equation}
where $\beta=0.0679$ and $\beta=0.222$ for explosive and random bond percolation. It is clear that the $\beta$ value for explosive is unusually smaller than that of its value for
random bond percolation. 

\subsection{Susceptibility: Mean cluster size S(p)}

\begin{figure}
\centering
\subfloat[]
{
\includegraphics[height=4.0 cm, width=2.4 cm, clip=true,angle=-90]
{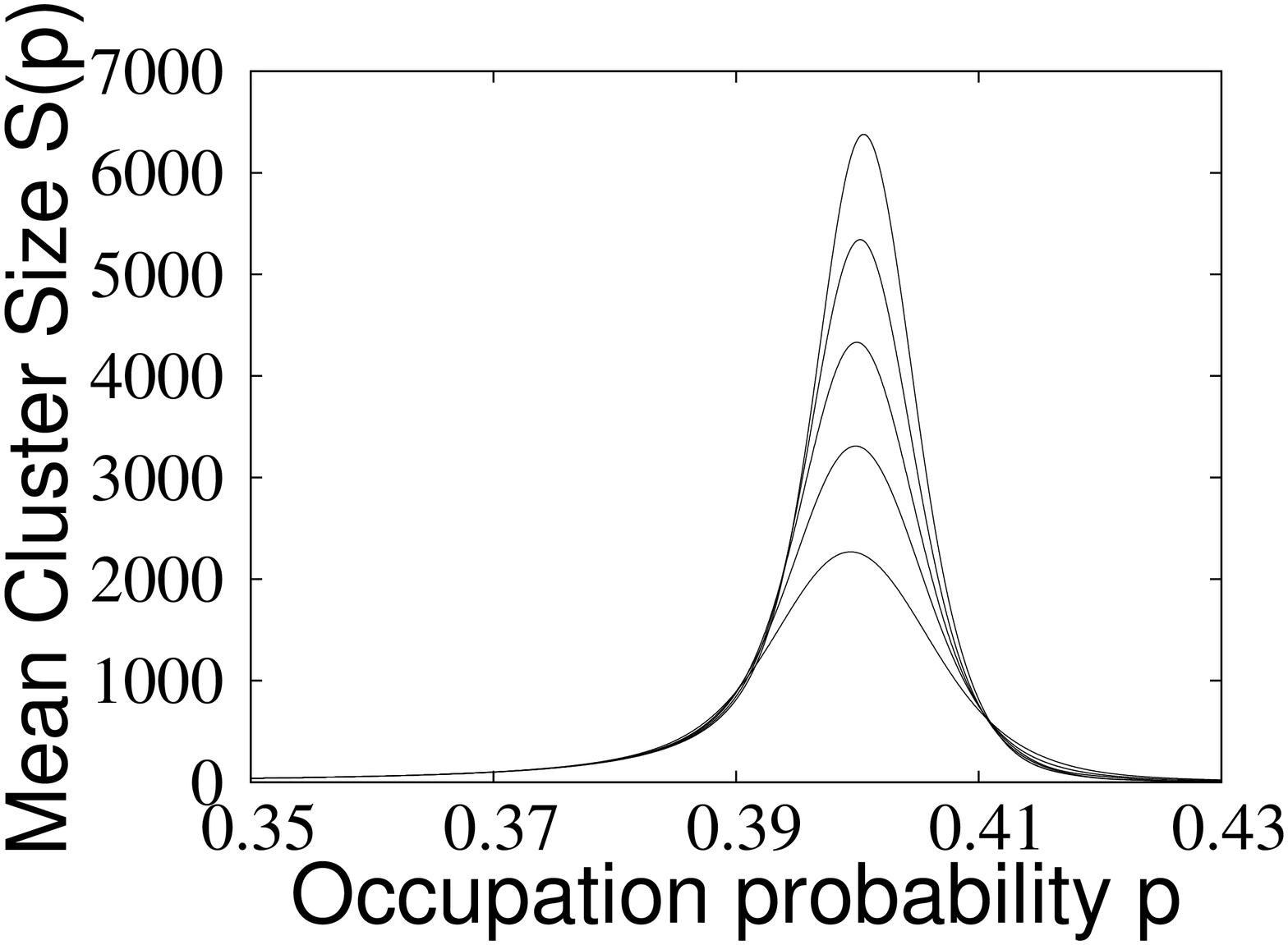}
\label{fig:4a}
}
\subfloat[]
{
\includegraphics[height=4.0 cm, width=2.4 cm, clip=true, angle=-90]
{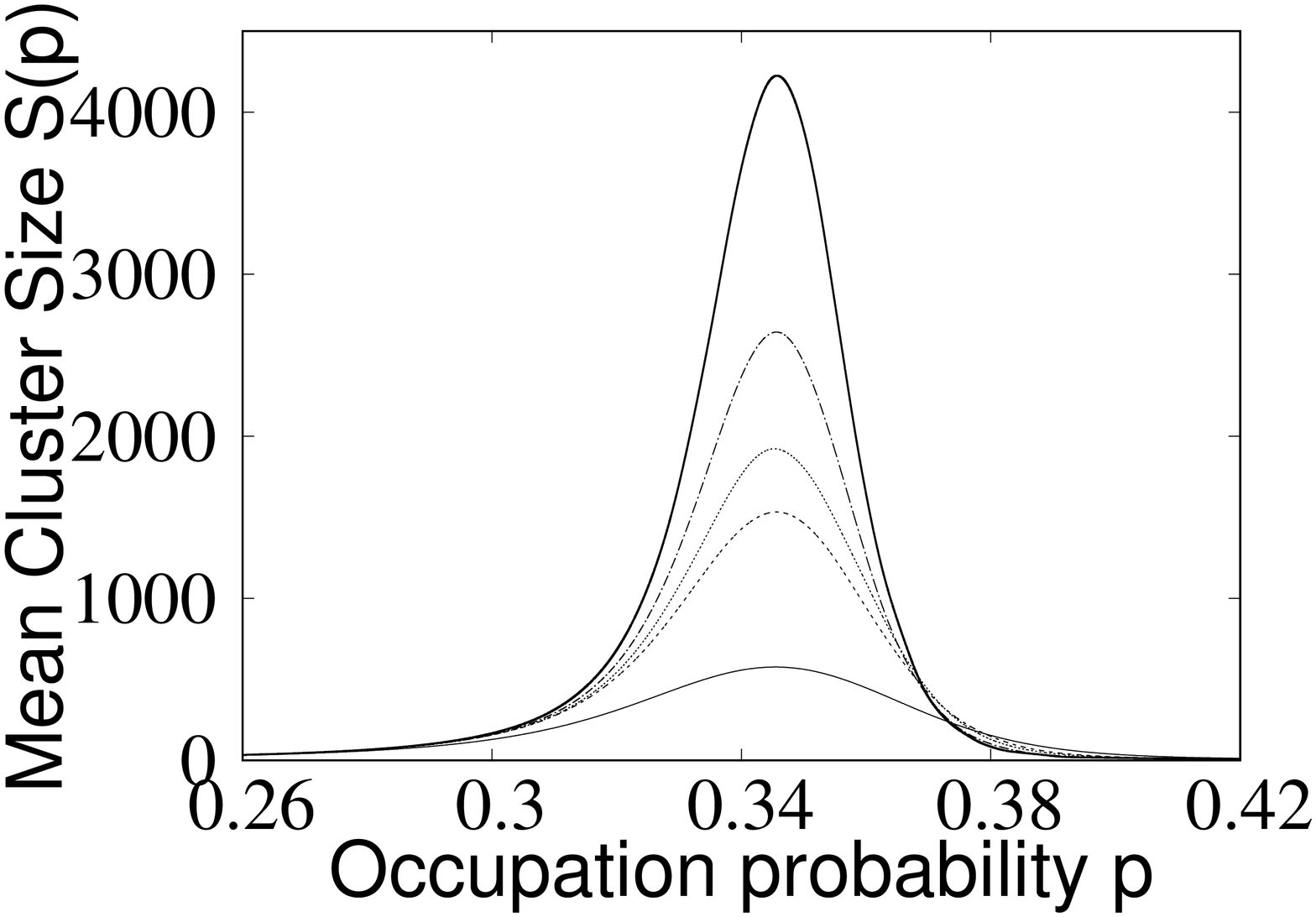}
\label{fig:4b}
}

\subfloat[]
{
\includegraphics[height=4.0 cm, width=2.4 cm, clip=true, angle=-90]
{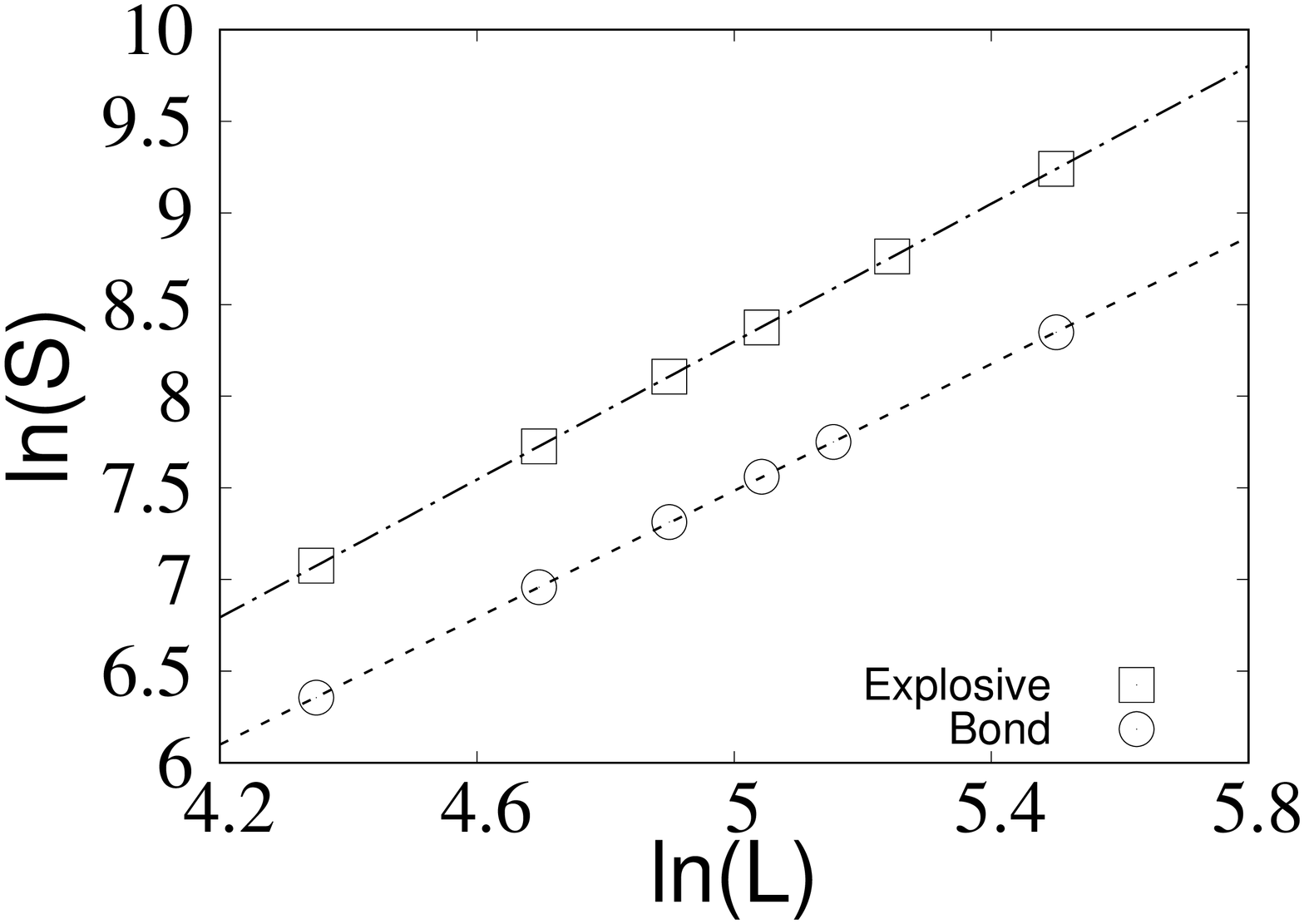}
\label{fig:4c}
}
\subfloat[]
{
\includegraphics[height=4.0 cm, width=2.4 cm, clip=true, angle=-90]
{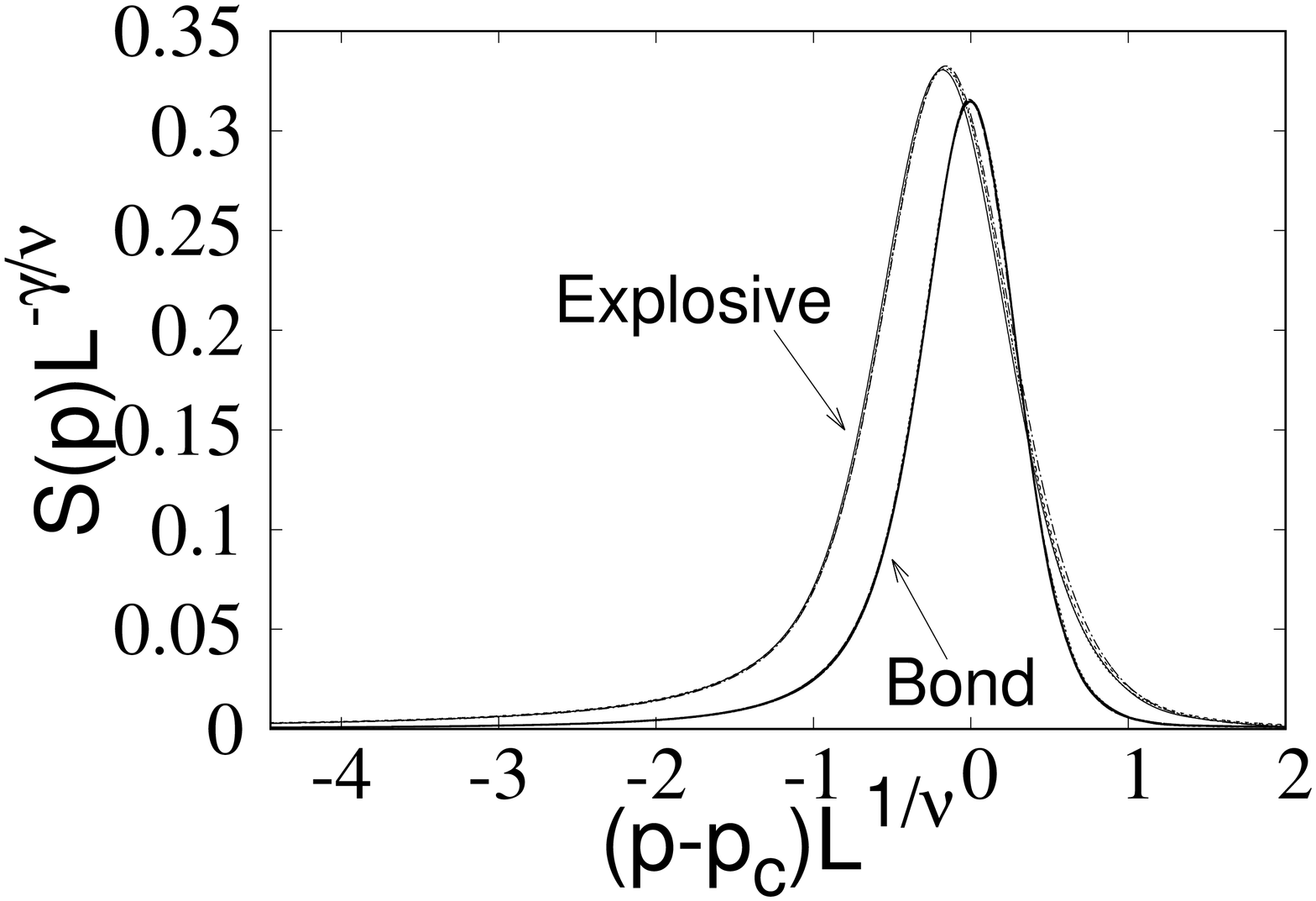}
\label{fig:4d}
}

\caption{The mean cluster size $S(p,L)$ for (a) explosive bond and (b) random bond percolation as a function of $p$ for different size of the WPSL. 
In the case of bond the cluster size is measured by the number of sites each cluster contains and
in the case of sites it is the area of the contiguous blocks that belongs to the same cluster. In (c) we plot $\log S$ vs $\log L$ using the size of $S$ for fixed value of $(p-p_c)L^{1/\nu}$ and find
almost parallel lines with slope  $\gamma/\nu$ equal to $1.8818 \pm 0.0069$ and $1.7315\pm -0.0019$ for explosive and random bond percolation respectively.
In order to obtain a better estimate for the $\gamma$ value we also plot the same data of (a) and (b) in the self-similar coordinates namely
$C_2SL^{-\gamma/\nu}$ vs $(p-p_c)L^{1/\nu}$ in (d). We again find that all distinct plots of (a) and (b) 
collapse into their respective universal curve.
} 
\label{fig:4abcd}
\end{figure}

The mean cluster size is regarded as the equivalent counterpart of the susceptibility. 
Using the idea of the cluster size distribution function $n_s(p)$,
the number of clusters of size $s$ per site, we can define the mean cluster size $S(p)$ as
\begin{equation}
\label{eq:nsp3}
S(p)=\sum_s sf_s={{\sum_s s^2n_s}\over{\sum_s sn_s}},
\end{equation}
where the sum is over the finite clusters only i.e., the spanning cluster is excluded from the enumeration
of $S$. In Figs. \ref{fig:4a} and \ref{fig:4b} we show the plots of $S(p)$, for both 
explosive and random bond percolation, as a function of $p$ 
for  different lattice sizes $L=\sqrt{N}$. We observe that in either cases, the peak height 
grows profoundly with $L$ in the vicinity of $p_c$. To find the critical
exponent $\gamma$ we first plot $S$ vs $(p_c-p)L^{1/\nu}$ and find that the peak heights $S_{{\rm peak}}$
lie along 
the same line. We then measure the size of the $S_{{\rm peak}}$ for different $L$. 
Plotting $\log[S_{{\rm peak}}]$ vs $\log(L)$ in Fig. (\ref{fig:4c}) we find a straight line for both 
explosive and random bond percolation revealing that 
\begin{equation}
\label{eq:nsp5}
S_{{\rm peak}}\sim L^{\gamma/\nu},
\end{equation}
where we find that $\gamma/\nu$ equal to $1.8818 \pm 0.0069$ and  $1.7280 \pm 0.0019$ for 
random bond percolation. 
Plotting now $SL^{-\gamma/\nu}$ vs $(p_c-p)L^{1/\nu}$  in Fig. (\ref{fig:4d})
we find that all the distinct plots of Figs. (\ref{fig:4a}) and (\ref{fig:4b}) 
collapse superbly into universal curves. Such a data-collapse is a
clear testament that the mean cluster size too exhibits finite-size scaling 
\begin{equation}
S(p,L)\sim L^{\gamma/\nu}\phi_\gamma((p-p_c)L^{1/\nu}).
\end{equation} 
Eliminating $L$ from Eq. (\ref{eq:10}) in favor of $(p_c-p)$ using $(p_c-p)\sim L^{-1/\nu}$ we 
find that the mean cluster diverges 
\begin{equation}
\label{eq:nsp7}
S\sim (p_c-p)^{-\gamma},
\end{equation}
where $\gamma=2.13816$ and $\gamma=2.825$ for explosive and random bond percolation respectively. 
This value is significantly different from the 
known value $\gamma= 2.389$ for all the regular planar lattices.

\begin{figure}
\centering
\subfloat[]
{
\includegraphics[height=4.0 cm, width=2.4 cm, clip=true,angle=-90]
{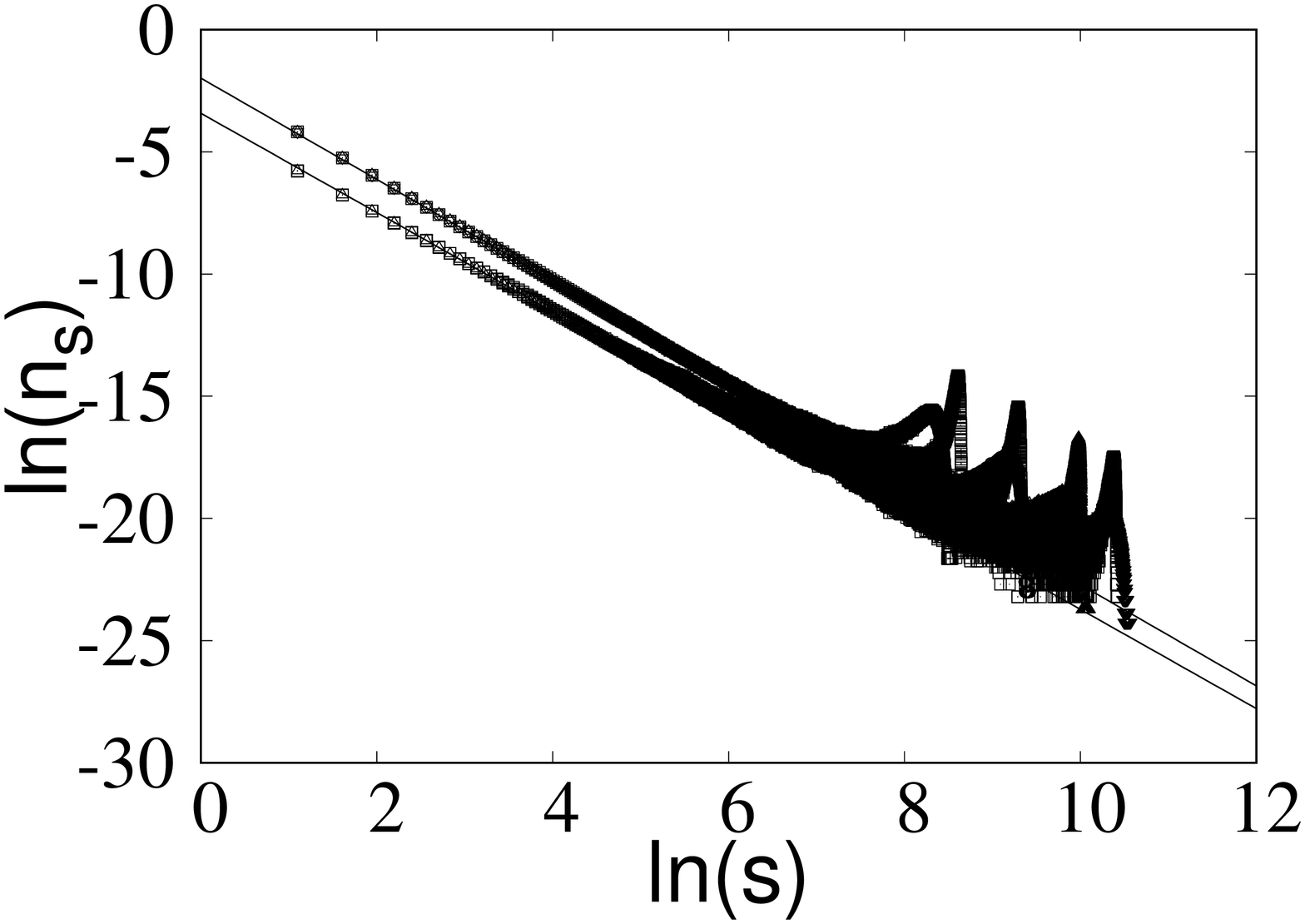}
\label{fig:5a}
}
\subfloat[]
{
\includegraphics[height=4.0 cm, width=2.4 cm, clip=true, angle=-90]
{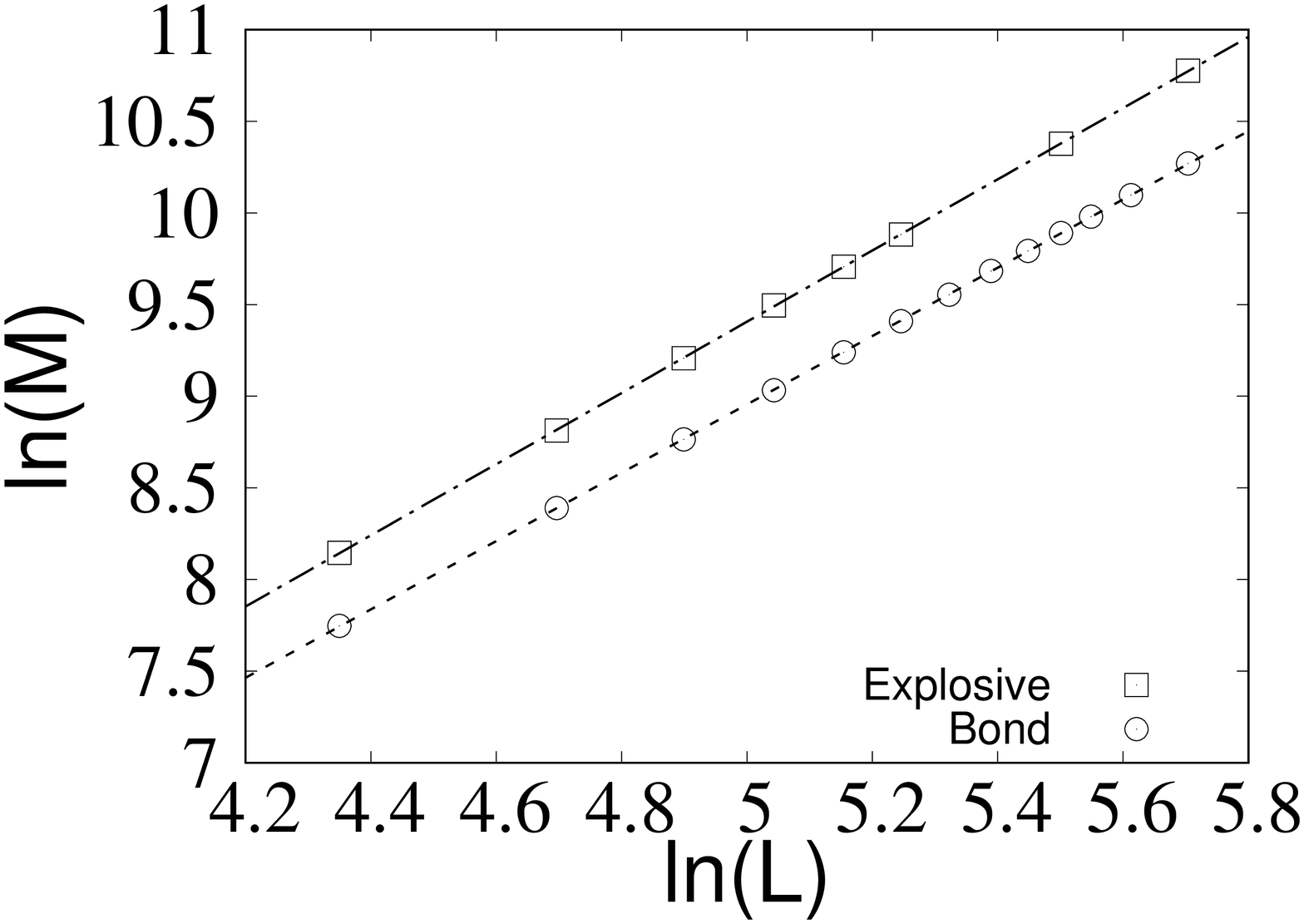}
\label{fig:5b}
}

\caption{We plot (a) the cluster size distribution function $\log(n_s(p_c))$ vs $\log s$ for different 
size of the WPSL and find almost parallel lines with slopes are $2.030$ and $2.0728$ for explosive and random
bond percolation respectively. It implies that the $\tau$ value is independent of the type of percolation.
(b) The mass of the spanning cluster $M$, the total area in the case of site and the number of sites in the case of bond, is shown as a function of system size $L$.
The two lines with slope $d_f=1.9415 \pm 0.0055$ for explosive and $1.8643 \pm 0.0014$ for bond once again reveals that the fractal dimension of the spanning cluster is independent of the type of percolation.
} 
\label{fig:5ab}
\end{figure}

It is well known that the cluster size distribution function $n_s(p)$  obeys 
\begin{equation}
\label{eq:nsp8}
n_s(p)\sim s^{-\tau}\phi((p-p_c)^{1/\sigma}s),
\end{equation}
and hence at $p=p_c$ it is 
\begin{equation}
\label{eq:tau}
n_s(p_c)\sim s^{-\tau}
\end{equation}
where $\tau$ is called the Fisher exponent.
To obtain the value of $\tau$ numerically we plot  $\log[n_s(p)]$ vs $\log(L)$ always at $p_c$
 in Fig. (\ref{fig:5a}) for both explosive and random bond percolation. The resulting plot in both the cases 
are straight lines with a hump near the tail 
due to finite size effect. However, we also observe that 
as the lattice size $L$ increases the extent up to which we obtain a straight line increases too. It implies
that if the size $L$ were infinitely large, we would have a perfect straight line obeying
Eq. (\ref{eq:tau}). The slopes of the lines are $\tau=2.030$ for explosive and  $\tau=2.0725$ for random 
bond percolation. It implies that the exponent $\tau$ is
almost the same $\tau\sim 2.072$ for both site and bond percolation on WPSL and its value is
different than the value for all known planar lattices $\tau=2.0549$.

\subsection{Fractal dimension of the spanning cluster}

Let $M(L)$ denote the mass or size of the percolating cluster at $p_c$ of linear size $L$.
Now we check the geometric nature of the spanning cluster.
First, if the cluster is an Euclidean object, then its mass $M(L)$ would grow  
as $M(L)\sim L^d$ with $d=2$ since the dimension of the embedding space of the WPSL is $d=2$. 
Now, a litmus test of whether the spanning cluster is a fractal or not would be to check if 
the exponent $d=2$ or $d<2$. If we find $d<2$ then that would mean the density of occupied sites is less as $L$ increases which would 
essentially mean that the spanning cluster is ramified or is stringy object.
To find the value of $d$ in the present case we plot the size or mass of the spanning cluster $M$ as a function of lattice size 
$L$ in the log-log scale as shown in Fig. (\ref{fig:5b}). Indeed, we find a
straight line with slope $d_f=1.9415\pm 0.0055$ for EBP and $d_f=1.8637 \pm -0.0224$ for RBP. The difference between the two values may appear small 
but it is important to remember that even a small difference in fractal dimension has a huge 
impact in its degree of ramification. It is well known that the numerical values of the various exponents $\beta, \gamma, \tau, d_f$ etc. for RBP cannot just
assume any arbitrary values rather they are bound by some scaling and hyperscaling relations. We find that the same is true also for EPB as
we find its exponents too are bound by the same scaling and hyperscaling relations such as 
$\tau=3-\gamma \sigma$, $\tau=1+d/d_f$, $\beta=\nu(d-d_f)$, $\gamma=\nu(2d_f-d)$ etc. We find that our estimates for various critical exponents satisfy these relations up to quite a good extent
regardless of whether it is about EBP or RBP.

\begin{table}[h!]
\centering
    \begin{tabular}{| l | l | l |}
    \hline
    Exponents & RBP on WPSL & EBP on WPSL \\ \hline
    $\nu$ & 1.635 & 1.136 \\ \hline
    $\beta$  & 0.222 & 0.0679 \\ \hline
    $\gamma$  & 2.825 & 2.137 \\ \hline
   $ \tau$  & 2.0728 & 2.03 \\ \hline
$d_f$ & 1.864 & 1.941 \\

    \hline
    \end{tabular}
\caption{The characteristic exponents for explosive and random bond percolation on the WPSL.}
\label{table:1}
\end{table}

\section{Summary and discussion}

In this article, we have studied explosive bond percolation on WPSL using extensive Monte Carlo simulations. 
The primary goal of this article is to study explosive bond on the WPSL. To this end, we have first obtained the
percolation threshold $p_c=0.4021$ for EBP which is greater  
than the $p_c=0.3457$ of the random bond percolation, as expected. We studied numerically the 
spanning probability $W(p)$, the percolation strength $P(p)$ and the mean cluster size $S(p)$
using the NZ algorithm. The resulting data is then used in the finite-size scaling theory to
obtain the various critical exponents $\nu, \beta$, $\gamma$ as well as other related exponents like $\tau$ and $d_f$. To that end, we obtained them numerically for EBP and 
compared them with those for the RBP 
 (see table \ref{table:1} for detailed comparison). Note that in all cases we
 found excellent data collapse. The quality of data-collapses provide a clear testament that the 
estimated values for various exponents are exceedingly close to the exact value. Besides, we found that these
values obey all the scaling and hyperscaling relations like we find in the random percolation. It implies
that EP is no special except the fact that the $\beta$ value is extremely low compare to the value we
find in the random percolation. Such low $\beta$ value makes it difficult to distinguish the behavior of the 
order parameter whether
it has really suffer a jump or show continuity. This was exactly the reason why explosive percolation was considered to describe 
first order transition.

Note that a comprehensive study of EBP to find critical exponents and to classify it into universality classes
has not yet even begun. In contrast, the classification of the random percolation is extensively studied 
and the results are quite interesting. For instance, 
it has been found that the numerical values of the critical exponents 
are universal in the sense that their values depend only on the dimension of the lattice. Their values neither 
depend on the detailed nature of the
structure of the lattice nor on the type of percolation i.e., whether the percolation is site or bond type. 
Remarkably, similar classification has also been found true in the case of models for thermal
continuous phase transition. Indeed, it has been found that the corresponding critical exponents of 
thermal phase transition neither depend on the lattice structure nor on the nature of interaction, but only on 
the spatial dimensionality, spin dimensionality and the range of the interactions. Recently, we have shown that 
random percolation on WPSL does not belong to the same
universality class where all the known planar lattices belong despite the dimension of the WPSL and that of 
the space, where WPSL is embedded, are the same. This
is really an exceptional case which is not so surprising owing to the fact that WPSL is itself an exceptional 
lattice.  We  hope that our findings will have a significant impact in the future study of the percolation theory 
especially in classifying the explosive percolation into universality
classes.

\end{document}